# Schlieren and BOS velocimetry of a round turbulent helium jet in air

Gary S. Settles[a] and Alex Liberzon[b]

[a]Corresponding author, 799 Walnut Spring Lane, State College, Pennsylvania 16801 USA, gss2@psu.edu
[b]School of Mechanical Engineering, Tel-Aviv University, Tel-Aviv, 6997801, Israel

ABSTRACT

Seedless velocimetry is gaining interest in many industrial and research applications. We report on a comparative study of time-resolved optical velocimetry using traditional, mirror-type knife-edge schlieren optics versus Background-Oriented Schlieren (BOS) of subsonic round turbulent helium jets in air at $Re_d$ = 5,890 and 11,300. Digital images with 1024 pixels streamwise resolution ($0 \leq x/d \leq 200$) were captured at 6000 frames/s in large ensembles. Velocimetry was performed on these results by digital image correlation (DIC) using OpenPIV software, and by streak-schlieren analysis of $x$-$t$ diagrams (kymography). Limited PIV data were also collected for verification of the schlieren velocimetry results. Both BOS and traditional schlieren show partial success in measuring the mean-flow helium jet self-similarity in terms of the $1/x$ decay of centerline velocity, Gaussian-shaped radial velocity profiles, and linear spreading rate of the jet. Visualized turbulent eddies, used as tracers in schlieren velocimetry, are observed to last longer than is necessary for this purpose in the present helium jets. Also, the measured convective velocity appears to be sufficiently robust to sum to the jet mean velocity in some of the results. Kymography yields better overall results than DIC, which we attribute to kymography's spatiotemporal "spectrum" of jet velocities, enabling the discrimination of fast eddies near the jet centerline from slower ones near the jet periphery. DIC and other analysis methods suffer from a path-averaging bias which negatively affects the results. The reduction of kymographic data for velocimetry was done manually and also by a Fourier-transform image-feature-orientation code, both yielding equivalent results.

## 1. Introduction

Standard laboratory schlieren optics based on lenses or mirrors [1] are limited to imaging solids, liquid or gas mixing, thermal flows, or turbulent compressible flows. Formerly used mostly for qualitative visualization, schlieren instruments now include high-speed digital cameras and sophisticated image processing that yield quantitative data [2] such as wave tracking, frequency spectra, and modal decomposition. They also supply big data to train machine learning and infer 3D flowfields using physics-informed neural networks [3].

  Seedless optical velocimetry is one of these features of modern digital schlieren imaging. It was invented in 1936 by Townend [4], far ahead of the computers that were needed to power it. Townend's velocimetry, however, tracked only artificial schlieren tracers that he introduced into the flow. It is the antecedent of later attempts to tag flows with laser-induced hot spots that were then imaged by schlieren optics [5, 6].

Papamoschou [7] tracked eddies in high-speed schlieren records of compressible turbulent mixing layers in 1989 using a simple two-image scheme and manual data reduction. This important step used the turbulent structures themselves as Lagrangian velocity tracers. Garg and Settles [8] then used two-point correlations of focused-schlieren image intensity to measure the convective velocity in a supersonic turbulent boundary layer.

Optical Flow (OF) was first used for schlieren velocimetry by Strickland and Sweeny [9] in their 1988 combustion experiments. They may also have been first to construct streak images ($x$-$t$ diagrams or kymograms) from high-speed shadowgram sequences. Liu et al. derived a rational foundation for applying OF to flow visualization data [10], produced OpenOpticalFlow freeware to do this [11], and showed results of its application to schlieren image sequences [12]. Since then, a number of investigators have adopted an OF approach to schlieren velocimetry.

Roosen and Meisner [13] were the first to apply a Digital Image Correlation (DIC) to schlieren image sequences in a fascinating 1999 astronomy experiment, where a 0.75 m-aperture telescope was adapted to perform schlieren velocimetry of atmospheric disturbances by starlight. In 2006, Jonassen et al. [14] used PIV laser equipment and software of the day for DIC velocimetry of a Round Turbulent Jet (RTJ) and a supersonic planar 2D turbulent boundary layer. Both schlieren and shadowgraph image sequences were studied. Their term "Schlieren PIV" has since been replaced by "schlieren velocimetry." Given the ubiquity of modern PIV equipment, software, and high-speed cameras, the DIC approach to schlieren velocimetry has been used often over the last 15 years, e.g. Schmidt et al. [15].

Kouchi et al. [16] made an important contribution to schlieren velocimetry by measuring the convective speed of turbulent eddies in a high-speed turbulent boundary layer and a transverse jet using $x$-$t$ diagrams, a form of kymography. They extracted these data from the x-t diagrams by Fourier transform analysis.

The importance of statistically-significant image ensembles to schlieren velocimetry was emphasized by Hargather et al. [17], whose imaging speed and test time were inadequate to resolve the intermittent outer region of a 2D high-speed turbulent boundary layer using DIC. This problem was solved by use of a modern high-speed digital camera (Wills et al. [18]), who acquired an image ensemble of sufficient size to successfully measure the mean velocity of a similar 2D turbulent boundary layer all the way out to the freestream.

While most of the studies cited above involve laboratory schlieren instruments, the relatively-new technique of Background-Oriented Schlieren (BOS) has also been used for schlieren velocimetry. Some applications, especially those outside the lab, are unsuitable for large and expensive schlieren mirrors, but BOS is simple and inexpensive enough to serve, e.g. [19]. It can thus bring schlieren velocimetry to a broader audience and a broader range of flows. Published BOS velocimetry papers include Mittelstaedt et al. [20], who examined a hydrothermal vent on the ocean floor but could not stop the flow to take a background reference image. Bühlmann et al. [21] applied BOS with two-step DIC processing of data from a fire in a tunnel. Raffel described velocimetry in his 2015 BOS review [22] but showed no results. Weilenmann et al. [23] applied Optical Flow BOS to a 3D unsteady combustor flowfield, and finally Taberlet [24] demonstrated BOS streak-schlieren velocimetry (kymography) of the thermal plume of a cigarette lighter. Of these studies, none provides a quantitative demonstration using a known flowfield as a benchmark for verification. Such a benchmark study is needed in order to further advance the topic of BOS velocimetry.

Finally, several common limitations or concerns have been raised about the overall concept of schlieren velocimetry:
1. *Schlieren images are line-integrated along the optical path, so how can schlieren velocimetry work for 3D flows?* (Clearly it cannot, except in the context of tomography. It can succeed for planar 2D flows and, as addressed in this paper, axisymmetric 2D flows.)
2. *Schlieren velocimetry measures the convective speed of turbulent eddies, but that is not necessarily the same as a measurement of the mean flow speed.* (There is evidence that it is the same, at least under certain conditions: [17,18], but additional evidence is needed.)
3. *Despite Taylors' frozen turbulence hypothesis, turbulence does evolve with time, so it cannot be a useful Lagrangian tracer if $\Delta t$ between schlieren images is too large.* (Yes, but better knowledge is needed of the minimum schlieren frame rate required in order to avoid this problem in various flowfields.)

These and similar concerns are addressed in our goals for the present study.

## 2. Goals

This research strives to demonstrate that schlieren velocimetry, which works without particles, is useful and achievable while being non-intrusive, and that it can provide dense data with modest cost and equipment in some cases. To that end we attempt to:
- Carry out a schlieren velocimetry experiment using a round turbulent jet having a well-known self-similar mean flowfield, with which the degree of success can be directly quantified.
- Perform mean-flow velocimetry of large image ensembles by 1-step DIC of sequential mirror-schlieren images, by 2-step DIC of BOS images, and by streak-schlieren or kymographic analysis.
- Benchmark the schlieren velocimetry results by comparison with PIV data.
- Compare these velocimetry results with trends expected from RTJ self-similarity.
- Gain further insight into the limitations of schlieren velocimetry as listed above.

## 3. Coordinate Frame, Experimental Apparatus, and Procedures

### 3.1 Coordinate Frame

We rely primarily upon the well-known and well-documented self-similarity of round turbulent jets, both in single-fluid and two-fluid cases, as thoroughly described e.g. by Pope [25]. The RTJ, a classical 2-D axisymmetric turbulent flow, is assumed to be self-similar in a polar-cylindrical coordinate system with axial velocity $U_0(x) \sim 1/x$, where $x$ is the jet centerline coordinate. It is also assumed to have self-similar Gaussian radial velocity profiles and a linear jet spreading rate. The present experimental data are to be checked against these assumptions.

As shown in Fig. 1, $x$ is normalized by the nozzle diameter $d$. Three downstream locations, $x/d$ = 49, 114, and 181, are chosen for PIV measurements and for other uses to be described. The jet spreading rate, per [25], is defined as $S = \mathrm{d}(r_{1/2}(x))/\mathrm{d}x$, where $r_{1/2}(x)$ is the jet half-width.

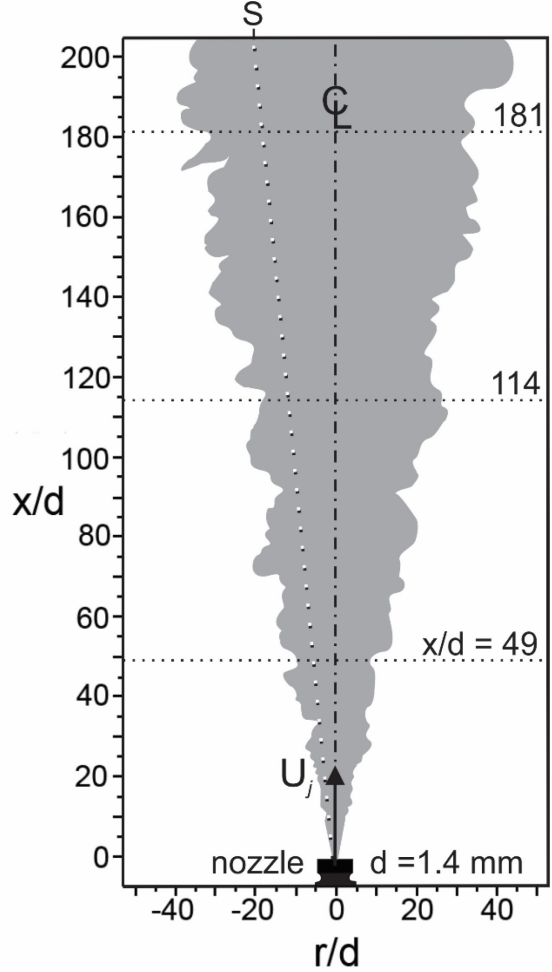

Fig. 1 Helium jet coordinate frame and parameters

*3.2 Helium Jet Equipment and Description*

Helium was chosen as a test gas primarily for its high refractivity, which is important to the sensitivity and resolution of the BOS technique. Pure helium gas was provided by a commercial cylinder (Airgas USA LLC) and a regulator connected to a chamber feeding the jet nozzle. Gas stagnation pressure $p_0$ and temperature $T_0$ were read from the chamber by a calibrated reference gage (Wallace & Tiernan) and a thermocouple. $T_0$ was typically ~ 285 K across the test series. The nozzle included a convergent section followed by a round, sharp-edged, straight orifice of 2.2 mm length and $d$ =1.4 mm diameter.

    The jet exit velocity $U_j$ is also a key similarity parameter that was not measured, but rather was calculated according to isentropic gas dynamics theory, e.g. [26]. The jet pressure ratio, $p_0/p_a$, where subscript a denotes atmospheric conditions, and the helium ratio of specific heats, $\gamma =1.667$, yield the jet exit Mach number $M_j$, whence $U_j$ is obtained using the definition of acoustic speed and the helium gas constant, $R = 2077.2$ Joule/(kg K). Finally $Re_d$, the jet Reynolds number based on $d$, is found using temperature-viscosity data for helium [27]. Two helium stagnation pressures were chosen: $p_0 = 119$ and 266 kPa, yielding $M_j = 0.45$ and 0.75, $U_j = 436$ and 682 m/s, and $Re_d = $ 5,890 and 11,300, respectively. While the choice of helium provides the refractivity needed for

schlieren imaging, the compressibility introduced by $M_j$ up to 0.75 is a necessary complication in order to achieve turbulent Reynolds numbers and fully-developed RTJs within our limited optical field-of-view.

Although there are many fine RTJ experiments in the literature, relatively few involve helium jets in air. Panchapakesan and Lumley [28] made an important contribution in 1993, in which they took special note of the effect of the air-helium density ratio $\omega = \rho_a/\rho_h = 7.24$ and the densimetric Froude number, $F = U_j^2 \rho_j/(\rho_a - \rho_j)gd$, which determines the importance of buoyancy. The present helium jets have $F = 2.4\times10^6$ and $6.8\times10^6$ for the low and high-$Re_d$ cases, respectively, thus both are completely in the Non-Buoyant Jet regime. Self-similar solutions of the equations of fluid motion are expected in this regime except where strong density fluctuations occur, i.e. for $x/d$ less than about 50 [28].

### 3.3 Schlieren Apparatus

As shown in Fig. 2, two separate optical instruments were used in this study. The traditional mirror-type schlieren apparatus [1] (hereafter referred to as *mirror-schlieren*) was based on a 0.5 m diameter, f/4.2 spherical "sandwich" mirror made by hubbleoptics.com. This mirror allowed a schlieren field-of-view of the helium jet from the nozzle exit out to about $x/d = 200$, as shown in Fig. 1. The helium nozzle was located 0.25 m in front of the mirror. The mirror was illuminated at its radius of curvature by a battery-powered Cree XM-L2 continuous white-light LED, 10W at 3.7V, with a 2.1×2.1 mm emitter that was limited by an adjustable iris diaphragm. A dual 90° reflector, made from two small flat elliptical secondary telescope mirrors, directed the light beam to the mirror and intercepted its return reflection with a small horizontal parallax. A Photron Fastcam APX-RS digital CCD camera was used for all schlieren imaging. An AFS- Nikkor 70-300 mm zoom lens gave the desired image size at almost full zoom. The horizontal schlieren knife-edge was set to provide a 50% cutoff of the ~ 2 mm-diameter light source image. No photogrammetric correction was made for the slight lack of parallelism of the schlieren beam.

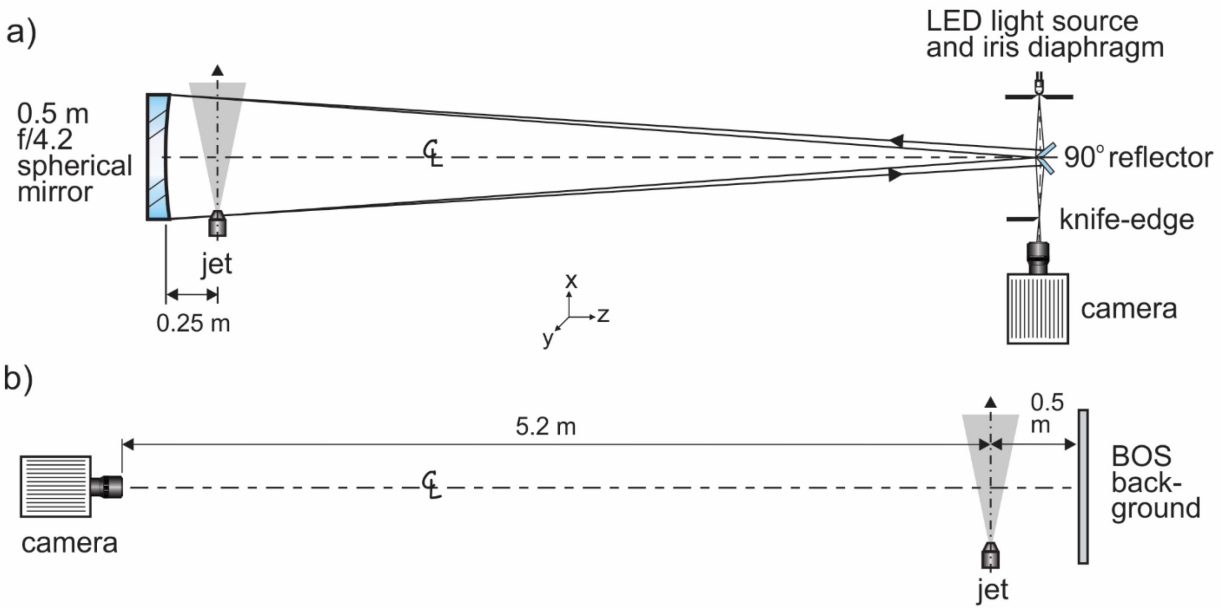

Fig. 2 a) traditional single-mirror schlieren apparatus, b) BOS apparatus.

The BOS setup was designed to provide the same jet coverage as the mirror-schlieren. Its length was limited to about 6 m by the available space. The BOS background was printed on a transparency, affixed to a window, and trans-illuminated by direct sunlight. The same camera, lens, frame rate, and field-of-view served both the BOS and the mirror-schlieren instruments. However, the key compromise of BOS imaging requires a large standoff distance of the jet from the background for high sensitivity, or else a small standoff distance for reasonably-sharp image focus. We opted for the latter, making the standoff distance as small as possible while still maintaining enough schlieren sensitivity to properly image the helium jet. That led to the 0.5 m distance from jet to background shown in Fig. 2b. Even so, the BOS resolution could not approach that of the mirror-schlieren system, which was measured to be 1 line-pair per mm in the $x$-direction using a USAF 1951 test target. In these images, 1 image pixel = 0.29 mm.

Early BOS experiments with a random-dot background gave poor sensitivity, so a fine grid of horizontal lines was tried instead. Dalziel et al. [29] cite one background line for every two pixel rows on the camera sensor as the ideal, though perhaps-not-attainable spacing. This gave high streamwise sensitivity but also caused serious issues with moiré fringing in the BOS image and lacked any sensitivity in the $r$-direction. These issues were resolved by using a checkerboard grid made by Wildeman's [30] *genchkbd.m* MATLAB script, as shown in Fig. 3. The checkerboard squares typically occupy 2×2 pixels in the BOS image, giving 8 black squares or "dots" in an 8×8-pixel DIC interrogation window. Fig. 3 shows a segment of an actual raw BOS image illustrating this result. Since typical BOS deflections due to refraction in the helium jet are less than 0.4 pixel, there is no concern over aliasing errors due to this regular – rather than random – background.

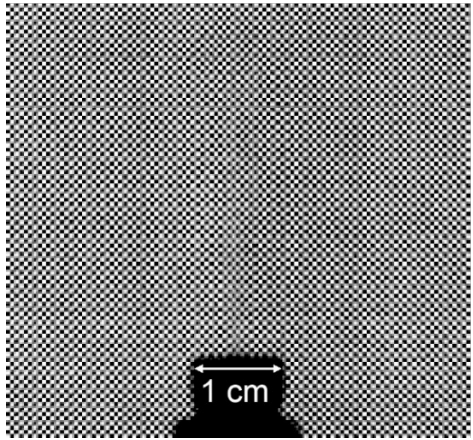

Fig. 3 Segment of a raw BOS image showing the helium nozzle and the checkerboard background. In the nozzle plane, 1 image pixel = 0.26 mm.

The sharp edge of the nozzle silhouette is blurred over a distance of 10 pixels (2.6 mm) in Fig. 3, solely due to defocusing the nozzle with respect to the sharply-focused BOS background, caused by the 0.5 m nozzle standoff distance. Upon DIC processing of BOS image pairs, additional blurring occurred due to the interrogation window size. The smallest interrogation window used in this study was 8×8 pixels, which is similar in size to the defocusing blur.

The Photron APX-RS camera provides 1024×512-pixel images at 6000 frames/s and 167 μs image exposure for the present schlieren imaging, and is able to record up to 2 s of real-time imagery at this rate (12,288 images). Higher camera frame rates are available at lower spatial resolution, but we opted instead to maintain high spatial resolution of the entire helium jet over $0 \leq x/d \leq 200$. The helium jet is very rapid in the nozzle exit near-field, so it is not time-resolvable at 6000 frames/s. However, further downstream, with jet velocities in the 3-30 m/s range, smearing of turbulent eddies should be limited to ½ to 5 mm, respectively. Thus we concentrate on $50 \leq x/d \leq 200$ and realize that the present schlieren velocimetry results will have the best accuracy toward the downstream end of the jet field-of-view.

*3.4 PIV*

Limited traditional PIV measurements were made in the jet centerplane in order to provide ground-truth data for comparison with schlieren velocimetry, especially in terms of $U_0(x)$. The Photron Fastcam APX-RS camera was again used for data acquisition, now fitted with a fixed Nikkor f/1.4 35 mm lens for close-up imaging. A 1W CW Diode-Pumped Solid-State (DPSS) Nd-YAG laser (Laserglow Technologies) provided sheet illumination at 532 nm. Since this laser was not powerful enough to illuminate the entire jet for high-speed PIV imaging, a narrow laser sheet of width 1 mm and height ~2 cm was generated using an $f = 100$ mm cylindrical lens. This laser sheet was precisely located on the vertical jet centerline using a Klein 93LCLS cross line plumb-spot laser level. PIV radial-profile measurements were made across the jet width at $x/d = 49$, 114, and 181 at both $Re_d = 5,890$ and 11,300. Image ensembles of up to 22,000 frames were recorded with 256×512 pixel size, 15,000 frames/s speed and 25 μs exposure at $x/d = 181$, and up to 30,000 frames/s and 12 μs exposure at $x/d = 49$. These settings were chosen based on observed frame-to-frame particle motion and in-frame particle smear.

Particle seeding was provided by a LeMaitre G100 theatrical fog generator, which is believed to generate ~1 μm-diameter particles [31]. The seeding was introduced via a manifold that produced a low-speed, seeded coaxial airflow at the nozzle for subsequent entrainment into the jet. The helium stream itself was not seeded prior to the nozzle exit.

*3.5 Schlieren Velocimetry Analysis*

To estimate the convergence rate of turbulence statistics in the schlieren image data, the grayscale intensity of a single pixel on the $Re_d = 5,890$ jet centerline at $x/d = 181$ was extracted with ImageJ software [32] using the *Image/Stacks/Plot Z-axis Profile* command. This yielded a time series of length 12,288 with a time step of 167 μs. This time series was autocorrelated using MATLAB's *autocorr* function and integrated, giving a value of 0.45 ms for the integral time scale τ of the jet (see [25], Eq. 3-139). With this knowledge, according to Papageorge and Sutton [33], an image ensemble of 0.45 s duration at 6000 frames/s is estimated to have a 3% error in its mean value, while our maximum 2 s sample size at this frame rate has only a 1% estimated error. At least a ½ s sample (3000 images) was thus used in all velocimetry analyses.

Example schlieren and BOS pseudo-schlieren images are shown in Fig. 4. For the mirror-schlieren cases at both Reynolds numbers, shown in Figs. 4a and b, consecutive images from the 6000 frame/s ensembles were paired for velocimetry analysis and processed by digital image correlation using a FFT-based cross-correlation method in the OpenPIV software package, producing a velocity-like representation of the jet as described further below.

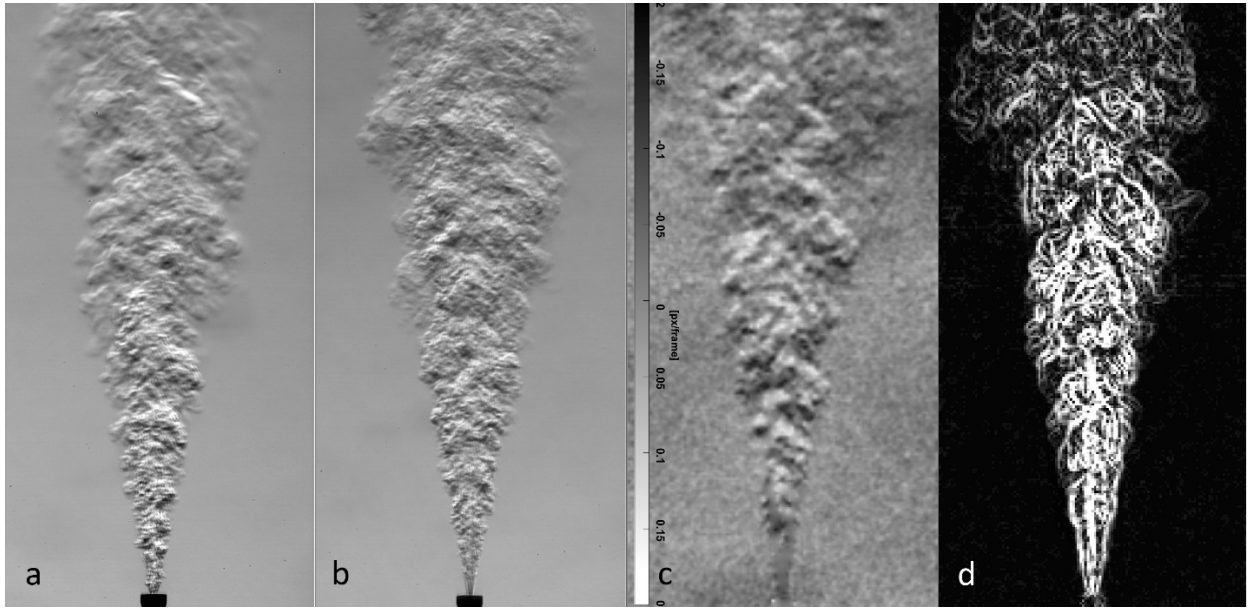

Fig. 4 Example schlieren and pseudo-schlieren helium jet images. a) horizontal-knife-edge mirror schlieren image at $Re_d$ = 5,890 and b) $Re_d$ = 11,300. c) BOS-DIC pseudo-schlieren image rendered with pseudo-horizontal-knife-edge cutoff at $Re_d$ = 5,890. d) BOS-strain-rate image at $Re_d$ = 5,890.

BOS-DIC data analysis was performed in three major steps: A) the FFT-based cross-correlation method in OpenPIV created a displacement field from each raw BOS image with respect to the no-flow reference (tare) image of the BOS background pattern (e.g. Fig. 3). B) These displacement fields were then post-processed to create pseudo-schlieren scalar fields of the streamwise pseudo-velocity component, velocity magnitude, velocity gradients, and strain-rate. An example BOS-DIC pseudo-schlieren scalar image is given in Fig. 4c, and a pseudo-strain-rate image in Fig. 4d. C) The pseudo-schlieren scalar fields are processed as raw image pairs by a second PIV run, creating a final velocity-like realization. The two PIV-like steps were performed with single-pass, multi-pass, and iterative-window-deformation multi-pass techniques (e.g. see [34]), and are available in the open-source Python version of OpenPIV [35]. For comparison, we also used the open-source PIVLab code [36, 37], and obtained essentially the same result.

   Steps A and B above produce pseudo-schlieren images that appear similar to mirror-schlieren images, but with less detail and more noise; e.g. compare Figs. 4a and 4c. Watching the animation of a BOS sequence clearly suggests the upward motion of the helium jet to an observer, but this does not always compare satisfactorily with the independent PIV data or with the mirror-schlieren velocimetry results. An attempt to generate high-pass-filtered data in the form of edges of displacement gradient fields (e.g. a strain-rate field), shown in Fig. 4d, also strongly suggests the upward jet motion qualitatively. Some other correlation methods were also attempted, including optical flow and multi-grid pyramids (filtering the data with filters varying from coarse-to-fine in size). The overall results of the BOS velocimetry processing are discussed in the next section.

Streak-schlieren analysis (kymography), as reviewed in the Introduction, was applied to both the mirror-schlieren and the BOS-DIC results using ImageJ software and the ImageJ plugin *KymoResliceWide* [38]. A schlieren image stack was opened in ImageJ and the *Straight Line* tool was used, with a linewidth of 1 pixel, to select the jet centerline as shown in Fig. 1. *KymoResliceWide* then extracted a single centerline pixel row (1×1024 pixels) from each image and stacked these rows to produce a new image that is an *x-t* diagram, streak diagram, or kymogram representing the motion of schlieren grayscale features in spacetime along the jet length and the duration of the image stack. Streaks in these schlieren kymograms represent the motion of turbulence as shown in the instantaneous images of Fig. 4. Moreover, the slopes of these streaks in spacetime represent the velocities of turbulent eddies and large-scale mixing structures. Velocimetry is done merely by measuring the streak slopes, as illustrated next.

Fig. 5 is a simplified diagram of a kymogram extracted from a high-speed helium-jet image sequence. The *x*-axis is the jet centerline *x*-axis (m), while the *t*-axis measures the time coordinate of the image sequence (s). Turbulent eddies in a round turbulent jet having $U \sim 1/x$ similarity follow curves like the red curve **a** and the blue curve **b** in this *x-t* diagram. Near $x = 0$, the nozzle exit location, the helium velocity is very high so the slope $dt/dx$ of the curve described by an eddy is quite small and the curve is almost horizontal. At large *x*, however, the jet has spread and slowed, causing turbulent streaks to curve concave upwards. Curve **b** portrays a streak from a turbulent structure that is moving slower than **a**, and thus has larger slopes $dt/dx$ at all *x* values. Any feature having zero velocity in this coordinate frame would create a vertical streak of infinite slope. In order to manually measure the streak velocity at the round yellow point shown on curve **a**, for example, one can manually fit a short line segment tangent to the curve at that point and measure d*t* and d*x* from it as shown in the figure. ImageJ's *Analyze>Measure* tool allows the cursor-drawing of such a line segment directly on a kymogram using the *Straight Line* tool, then outputs the measurement's *x*-centroid location, d*t*, and d*x* for further processing (in e.g. MATLAB or Microsoft Excel). The velocity at the measured point is simply $dx/dt$ in m/s. Such data are easy but time-consuming to collect manually using ImageJ, and there is good reason to seek automated data reduction of kymograms as discussed in the next section.

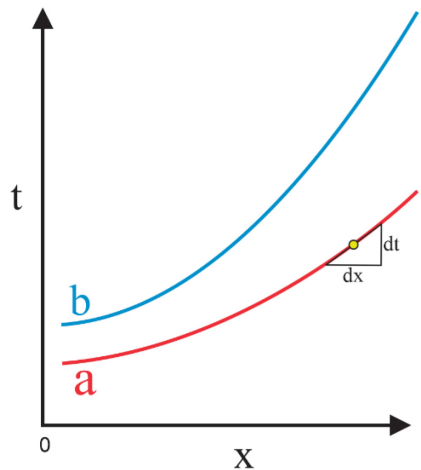

Fig. 5 Simplified drawing of a kymogram of turbulence streaks on the centerline of a high-speed jet

## 4. Results and Discussion

*4.1 PIV*

The 6 radial PIV jet profiles that were measured as described in Sect. 3.4 are all Gaussian in character. Their self-similarity scaling is demonstrated by Fig. 6, which is a graph of $U_j/U_0(x)$ vs. $x/d$ at $Re_d$ = 5,890 and 11,300 and $x/d$ = 49, 114, and 181. The error bars shown in the figure reflect the standard deviations of the data points as reported by PIVlab, and also that the high-$Re_d$ point at $x/d$ = 49 was approaching the limit of our measuring capability in terms of laser illumination. This graph demonstrates $1/x$ jet similarity within the experimental error, with no appreciable jet development region in the near-nozzle flowfield. The linear fit to the data is described by $U_j/U_0$ = $0.434(x/d)$ with $R^2$ = 0.9988. We note that Panchapakesan and Lumley [28] found the same equation for the centerline velocities of their helium jet, but with a slope of 0.414.

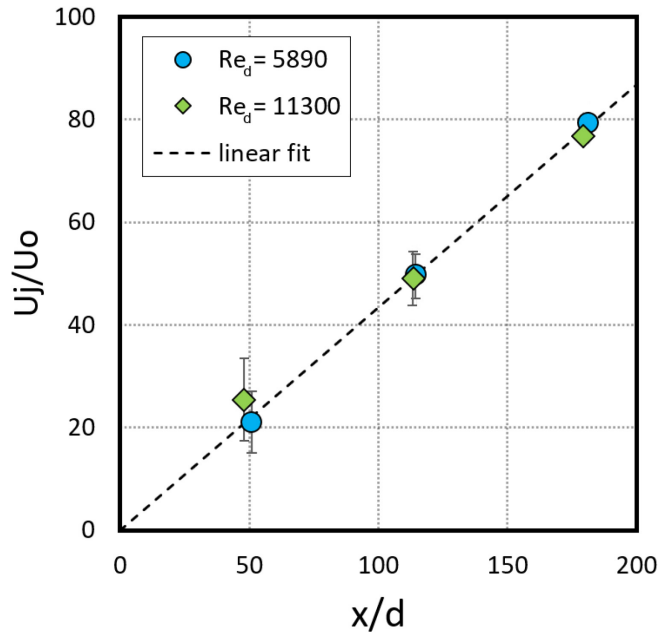

Fig. 6 Graph of PIV data in similarity coordinates.

*4.2 Schlieren Velocimetry by Digital Image Correlation*

Results from both BOS-DIC and mirror-schlieren DIC processing were considered in terms of their ability to yield velocities in accord with the established RTJ self-similarity rules. As a first example, radial jet velocity profiles $U/U_0$ vs. $r/r_{½}$ are shown in Fig. 7. Such a graph is often shown in both experimental [39] and computational [40] studies of RTJs in order to establish the validity of new results. Here, too, good agreement of radial profiles at three $x/d$ locations in Fig. 7 is evidence of the self-similarity of the jet velocity field and the ability of BOS-DIC-strain-rate image processing to measure it. Near the edge of the jet, however, the quality of the measurement fails because intermittency degrades the signal statistics. This is a problem for PIV as well, but schlieren velocimetry seems especially susceptible to it [17].

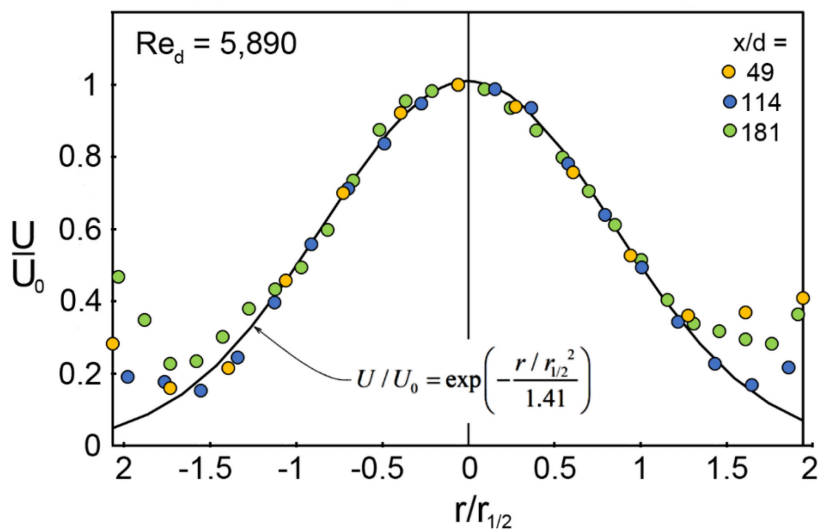

Fig. 7 Similarity graph of jet radial profiles $U/U_0$ vs. $r/r_{1/2}$ from BOS-DIC-strain-rate velocimetry at $Re_d = 5,890$.

Among the several approaches to the step-B processing of BOS displacement data that we tried, the simple grayscale rendering of pseudo-schlieren images, e.g. Fig. 4c, did not perform well in capturing smaller-scale, high-speed motion in DIC velocimetry step C. Gradient processing, which yields a strain-rate-like image such as Fig. 4d, was the most successful in producing jet centerline mean-velocity profiles vs. $x/d$ that are comparable to the PIV benchmark data. Such gradient processing simplifies a pseudo-schlieren image by edge detection, and discards some information about the brightness of large grayscale features. Jonassen et al. [14] similarly found that shadowgraph velocimetry of a helium RTJ in air outperformed schlieren velocimetry because the shadowgraph technique revealed the Laplacian of density fluctuations instead of the gradient. Schlieren images, especially with too much knife-edge cutoff, suffered clipping or over-ranging problems that led to large bright and dark grayscale features obscuring the underlying turbulent motion. (We did not include shadowgraphy here in order to limit the scope of the present study.)

Fig. 8 compares the $U_0(x)$ vs. $x/d$ behavior of the BOS-DIC-strain and mirror-schlieren data at $Re_d = 5,890$. In the BOS-DIC-strain case the indicated velocity rises rapidly to a peak with increasing distance from the nozzle exit, then less-rapidly declines. For $x/d < 20$ the jet was much too fast for the camera frame-rate, but the DIC was able to cope with this and eventually to reveal that $U_0(x)$ decreased with increasing $x$. Before $x/d$ reached 100, the BOS-DIC-strain data achieved a $1/x$-type velocity decay as shown by the red line in the figure. This indicates success in measuring a key feature of jet similarity. In the mirror-schlieren data, however, there is no pronounced velocity peak, but rather a plateau followed by a too-gradual decline. In this case the DIC is unable to recover from the insufficient frame rate at small $x/d$, and even continues to fail at large $x/d$ where the frame rate is adequate for DIC. In general, DIC velocimetry of mirror-schlieren images did not succeed for the present case of a helium RTJ in air.

This was unexpected, since the mirror-schlieren has much better resolution of turbulence than does the BOS pseudo-schlieren, as discussed earlier with regard to Fig. 4. However, it may be that the comparatively-low resolution of BOS discards only information that is superfluous for schlieren velocimetry in the present experiments.

Another possible contributor to the results of Fig. 8 is that the mirror-schlieren data pose a more severe challenge to DIC processing than do the BOS-DIC-strain data. The two images of a raw BOS image pair are very similar to one another, both closely resembling the example shown in Fig. 3 and having only sub-pixel shifts between them to be resolved by DIC. In the mirror-schlieren case, however, there is no underlying background pattern, so single-step DIC velocimetry must deal with frame-to-frame shifts of amorphous grayscale texture patches that may be from a few pixels to many pixels in size.

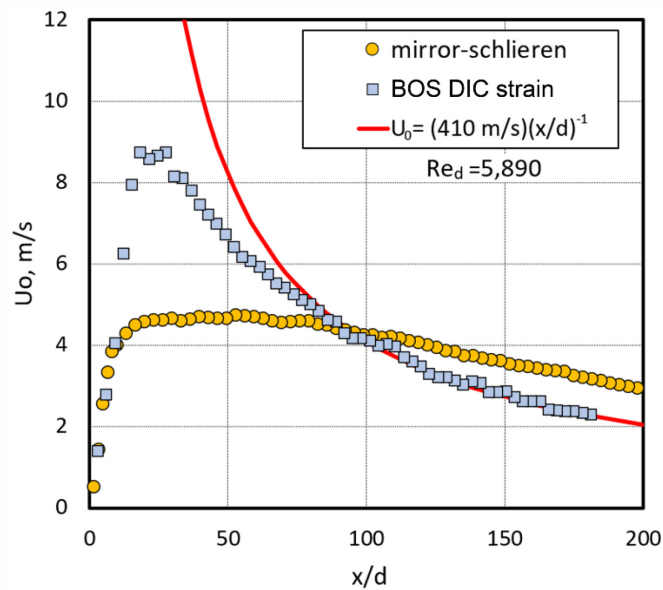

Fig. 8 Graph of $U_0(x)$ vs. $x/d$ velocimetry results of BOS-DIC-strain and mirror-schlieren DIC data at $Re_d$ = 5,890.

In contrast to this, DIC-based schlieren velocimetry was successful in measuring the mean velocity profiles of subsonic and supersonic 2D planar Turbulent Boundary Layers (TBLs) [14, 17] and a spatio-temporal DIC method by Wills et al. [18] also succeeded at measuring the mean-velocity profile of a compressible 2D planar TBL. Thus there is reason to attribute the poorer performance of DIC observed here to the nature of the axisymmetric jet, where all jet velocities from zero to $U_0$ are superimposed upon one another across the width of the jet by optical path-averaging, rather than being distributed along a single coordinate direction as they are in a 2D planar TBL.

Finally, we note that DIC is not an autonomous, trained machine intelligence that is put to work finding velocities from image data sets. It is "naïve" insofar as most of the above matters are concerned. It does not handle intermittent turbulence well, and it is not designed to account for the large variation with $x/d$ of the number of turbulent structures per interrogation window. However, it does fail benignly in the near-nozzle region and it recovers quickly downstream, at least for the present BOS-DIC-strain case. Also, as described next, naïve DIC is able to provide step-B-processed BOS pseudo-schlieren images that can be used to construct kymograms, which can then yield useful velocimetry results without requiring a second DIC step (step C).

*4.3 Schlieren Velocimetry by Streak-Schlieren (Kymography)*

An example helium-jet kymogram extracted from a mirror-schlieren image sequence at 6000 frames/s and $Re_d$ = 5,890 is shown in Fig. 9. Following the earlier description of Fig. 5, the $x/d$-axis is the jet centerline axis, while the $t$-axis measures the time coordinate of the sequence. (Only a small fraction of the 2 real-time seconds of data actually acquired is shown in this figure.)

The orange curve marked **a** in Fig. 9 is the schlieren track of a streak on the jet centerline, i.e., it is made by turbulent structures traveling at speed $U_0(x)$. Since centerline turbulence is the fastest turbulence to be found in the jet, a centerline streak has the shallowest slope to be found in the kymogram, which is a key to velocity data extraction. An off-centerline streak **b** has a similar shape but an overall higher slope distribution vs. $x/d$. Weak kymogram features such as **c** belong to low-speed turbulent structures near the edge of the jet. All these features are superimposed in the kymogram, but are nonetheless distinguishable mainly by their slopes.

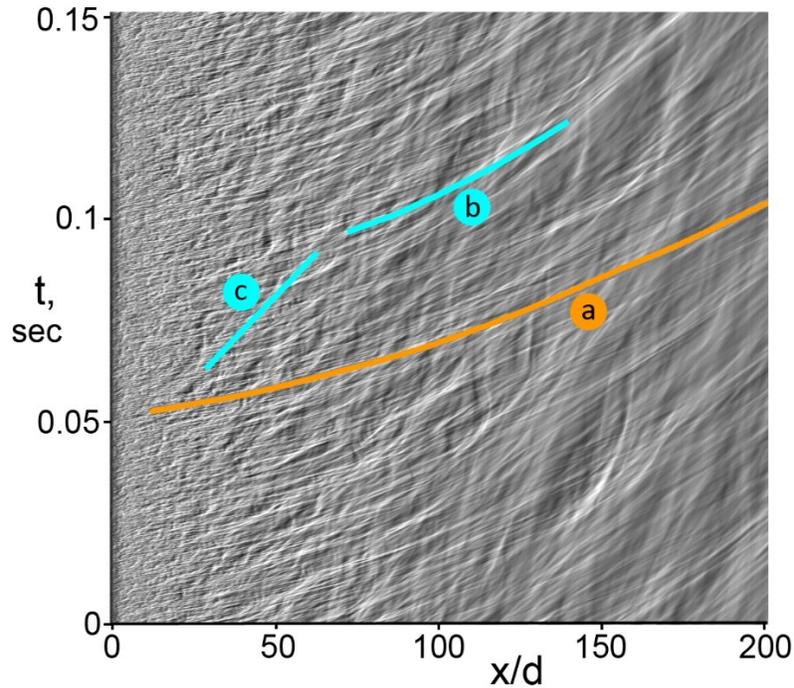

Fig. 9 Schlieren kymogram of jet turbulence in a centerline slice of the helium jet at $Re_d$ = 5,890.

To better understand how this jet kymogram was formed and what it represents, see Fig. 10a, a cross-sectional diagram of the schlieren light beam crossing the helium jet. The jet centerline or $x$-axis is perpendicular to the page and is shown by a red plus symbol in Fig. 10a. The schlieren beam enters from the left, as shown by light rays marked A, B, and C. Schlieren images, such as those of Fig. 4, are projected onto the camera sensor plane which is perpendicular to the page in this figure. The jet cross-section is portrayed in its radial plane by four "zones" of increasingly-darker yellow-green shading from the periphery to the center.

Rays A in Fig. 10a graze the edge of the jet and are affected only by a few slow, large-scale mixing structures found there. Rays B cross the jet midway from the edge to the centerline, passing through the $r_{½} = \pm 1$ points and being affected only by the first two of the four shaded regions in the cross-section. Ray C, however, passes through the jet centerline and is affected by all of the jet zones on its way in and out of the jet.

It is the nature of schlieren imaging that density-gradients are line-integrated across the jet by rays such as A, B, and C. However, the velocities of turbulent structures are not line-integrated, they are *path-averaged* (a crucial distinction that was recently emphasized by Schmidt et al. [16]). Thus ray C acquires velocity information across the entire jet but averages it, eventually yielding a velocimetry signal that must lie somewhere between zero velocity and the centerline velocity $U_0$ if subjected to DIC processing.

By constructing a kymogram such as Fig. 9, selecting only the centerline slices of each of a large sequence of jet schlieren images, we sample all of the turbulent structures affecting ray C from each of the zones depicted in Fig. 10a over the time duration of the sequence. This is a lot of information, even though only rays C are sampled and all other rays are ignored. If one wanted to sample all of the jet structures, multi-angle time-resolved tomography would be required. In this study, though, a single viewing angle, that of rays C in Fig. 10a, is sufficient to characterize all the typical turbulent motions in the jet by their velocities in the spatiotemporal "spectrum" produced by kymography.

To further emphasize this point, a different kymogram has been constructed by sampling a one-pixel-thick plane passing through the nozzle exit and a point midway between rows A and B, in other words a slice at an angle to the jet centerline and not far inside the jet edge. The resulting kymogram is shown in Fig. 10b.

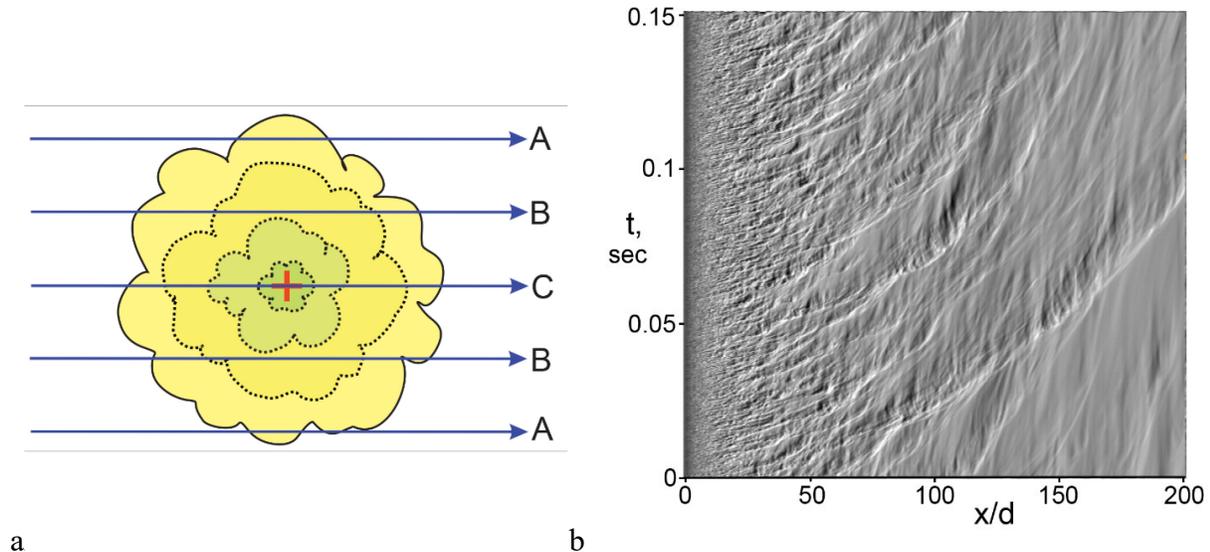

a    b

Fig. 10 a) diagram of the schlieren beam crossing the jet perpendicular to its axis. b) schlieren kymogram of turbulence in a slice of the helium jet near its edge, $Re_d = 5{,}890$.

Fig. 10b is similar to Fig. 9 in the nozzle near-field, where many rapid streaks are formed by small-scale turbulence, but for approximately $x/d > 90$ the only remaining kymogram features near the jet edge are large-scale organized mixing structures. An interesting comparison can be made between Fig. 10b and Figs. 4 and 6 of Mungal et al. [41], who show similar light-scattering $x$-$t$ diagrams of the periphery of a RTJ with a Reynolds number five orders of magnitude higher than ours.

Kymograms such as Figs. 9 and 10b are themselves a type of schlieren image in the spacetime domain. They have the appearance of a striated antique windowpane or of rough quarry stone illuminated by oblique sunlight.

Large image sequences, such as those generated here, can produce kymograms with thousands of streaks that are usable for velocimetry, thus providing considerable measurement redundancy. In addition to the mirror-schlieren kymograms shown here, BOS sequences processed as continuous-tone pseudo-schlieren images and as strain-rate images can also be used to construct usable kymograms, though at reduced resolution and, especially in the case of strain-rate images, at a lower signal-to-noise ratio.

### 4.4 Comparison of Jet Centerline PIV, DIC, and Kymograph Velocimetry Results

Fig. 11 shows the present schlieren velocimetry results in a similarity graph of $U_j/U_0(x)$ vs. $x/d$. The PIV benchmark is represented by its trendline from Fig. 6. Most of the present results lie above the PIV trendline, meaning that $U_0$ is underpredicted by schlieren velocimetry, as expected, due to path-averaging. However, the kymography results for mirror-schlieren and for BOS-DIC strain-rate processing at $Re_d = 11,300$ are in good agreement with the PIV trendline.

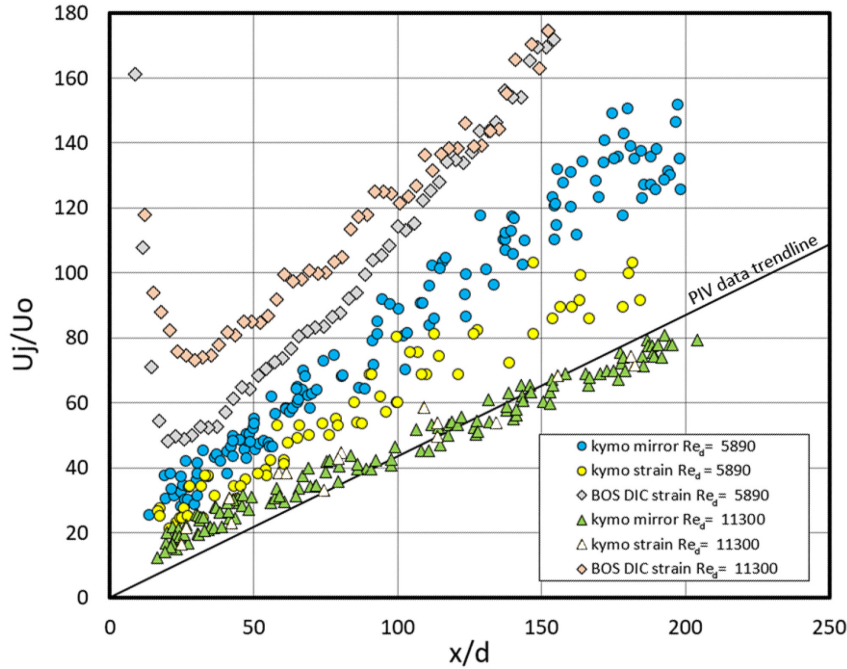

Fig. 11 Summary graph of present RTJ velocimetry results in similarity coordinates.

The kymography data, although having significant scatter bands due to the difficulty of measuring small slopes accurately, behave linearly in $U_j/U_0$ vs. $x/d$ coordinates in accordance with the self-similarity of RTJs. Curiously, these results all appear to originate from a common virtual origin at the unphysical value of $x/d = -33$. The BOS-DIC-strain-rate results also show approximate agreement with the kymography data at large $x/d$ in Fig. 11.

In order to better display and interpret these results in terms of schlieren velocimetry performance, the following Figs. 12 and 13 forego the $U_j$ nondimensionalization and instead portray dimensional $U_0(x)$ vs. $x/d$ for $Re_d = 5{,}890$ and $11{,}300$, respectively.

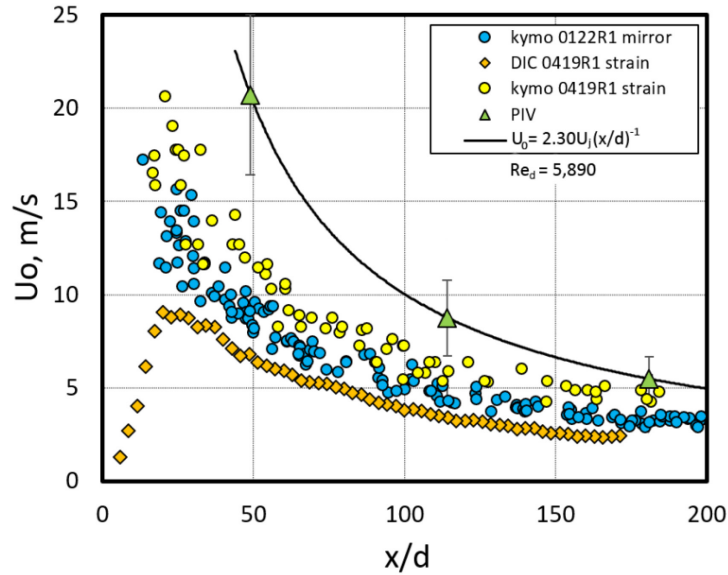

Fig. 12 Summary graph of $U_0(x)$ vs. $x/d$ for $Re_d = 5{,}890$.

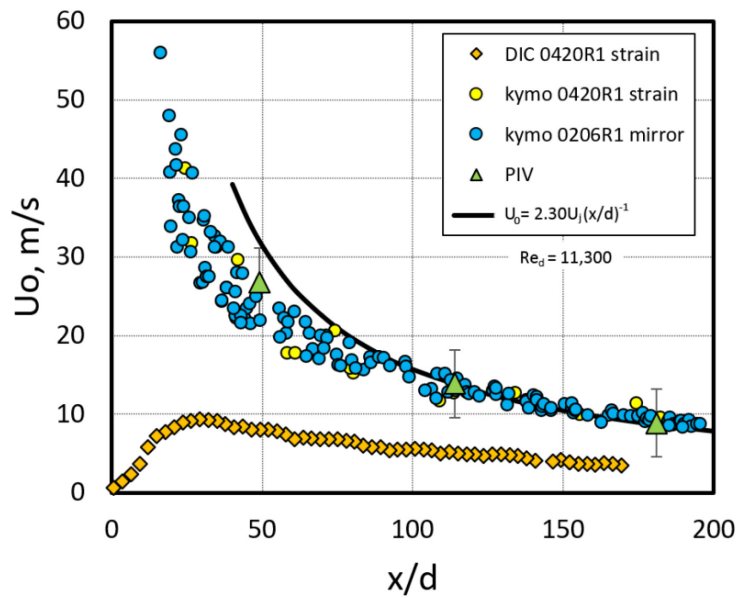

Fig. 13 Summary graph of $U_0(x)$ vs. $x/d$ for $Re_d = 11{,}300$.

In Fig. 12 it is clear that all present results for schlieren jet centerline velocity $U_0$ underestimate the PIV benchmark, although all do show an approximately $1/x$-type velocity decay with increasing $x/d$. This underestimate is expected due to path-averaging of the BOS-DIC-strain-rate data, but its cause is not so clear for the kymography data. The reason why kymography of the BOS-DIC-strain-rate results yields slightly-higher velocities than kymography of the high-resolution mirror-schlieren results is unknown. Between them, they indicate a jet centerline velocity of 50-60% of the PIV benchmark, while DIC processing of the BOS-strain-rate data predicts velocities only about 37% of the benchmark in Fig. 12.

The higher-Reynolds number case in Fig. 13 conveys a somewhat different picture. The kymographic results are now self-consistent and also in good agreement with the PIV benchmark for $x/d$ >100. Underpredicting the benchmark in the early jet region is not surprising, since our temporal data resolution there is admittedly poor and the most-shallow streak slopes in the kymograms are difficult to read accurately. But the agreement of kymography and PIV is a remarkable result: Kymography avoids path-averaging the jet velocity because it displays a spatiotemporal "spectrum" of the jet velocities, where fast turbulent structures near the jet centerline are distinguishable from slower ones near the jet periphery, and can be singled out by their shallow $x$-$t$ slopes. Kymography can therefore measure 100% of the jet centerline velocity at this Reynolds number for $x/d$ >100.

In contrast, the BOS-DIC-strain-rate data set in Fig. 13 is path-averaged, and it reports velocities only about 30% of the PIV benchmark velocity. The overall poor performance of DIC-based schlieren velocimetry in this study, especially for traditional mirror-schlieren-type input, is primarily attributed to the complexity of the axisymmetric RTJ compared to simpler, planar 2D test cases that have been tried previously. Nevertheless, our approach to the data analysis is neither exhaustive nor final. We have shared with the community our open-source code and some test cases with the hope of fostering better ideas for the future [42].

The Reynolds number effects in Figs. 12 and 13 are not well understood. The expected broadening of the turbulent length scale range with increasing $Re_d$ is seen qualitatively in Figs. 4a and 4b, for example, but quantitative data are needed. Pope's [25] overall description of RTJs requires that the jet and ambient fluids are the same in order to expect self-similarity, and then only when $Re_d > 10^4$. Our lower-$Re_d$ case shows ample evidence of self-similarity, but additional experiments well above $Re_d$ = 11,300 are needed in order to determine whether or not kymography's success, as shown in Fig. 13, is indeed a generally-applicable high-$Re_d$ RTJ attribute.

Finally, for the cases in Figs. 12 and 13 where schlieren path-averaging applies, there is the question: What fraction of the true $U_0(x)$ should we expect to recover from path-averaging on the jet centerline? General observations, e.g. [2], claim that it should be "about 50%." Schmidt et al. [16] recovered 64% of their PIV benchmark $U_0$ velocity in a particular jet example. If schlieren-velocimetry path-averaging were as simple as taking the arithmetic mean of a Gaussian jet velocity distribution along line C in Fig. 10a between the limits of $-2 < r/r_{1/2} \leq 2$, then it would yield 34% of $U_0$. The BOS-DIC-strain result of Figs. 12 and 13 do fall in this range, but one cannot assume such simplicity for turbulent-flow path-averaging, and further effort is needed to address this question for the future of schlieren velocimetry.

*4.5 Other Results*

Most kymography today is done in life-sciences research, and most of it is done manually according to [43]. Several applications and plugins, e.g. for ImageJ, are designed to aid this process while still requiring a human operator to select which kymogram tracks/streaks to digitize. Despite these aids, the manual reduction of large kymography datasets, like those produced here, can be very laborious, and there remains the possibility of operator subjectivity influencing the outcome.

The few available fully-automated approaches to kymography include FFT image processing [16], computer-vision line detection [44, 45], and machine-learning neural network codes [43, 46]. For present purposes we have chosen an FFT approach that is readily available as an ImageJ plugin: H. Glünder's *Orientations* 1.1.x [47, 48]. This plugin operates on an 8-bit monochrome image using overlapping disc-shaped windows of user-selected pixel size, and returns an analysis of local dominant feature orientations in each window. These results are given as orientation indicators overlaid on the input image, a histogram of feature orientations, and numerical output suitable for further analysis in MATLAB, Microsoft Excel, etc. This numerical output lists window number, window center location in x and t, and local feature angle in degrees.

We applied *Orientations* 1.1.x in ImageJ to automatically analyze kymograms such as Fig. 9, which were originally reduced manually to yield the kymography data given in Figs. 11-13. As before, the jet centerline velocity $U_0(x)$ was assumed to be given by the shallowest slopes, or fastest eddy speeds, at each $x/d$ location. The results are compared in Fig. 14 for mirror-schlieren and BOS-strain kymography at $Re_d$ = 5,890, and for mirror-schlieren kymography at $Re_d$ = 11,300. In each case the manual and automated data-reduction results are overlaid for comparison.

The upshot of Fig. 14 is that, within the respective scatter bands of the three cases considered, the manual and automated kymography data are essentially identical. This is an important result which supports the validity of kymography as a tool for quantitative schlieren velocimetry. We note, however, some key differences in the breadth of and effort required for manual vs. automated kymogram analysis: Manual analysis can be tedious, requiring perhaps 1/2 hour or more to acquire 150 $U_0(x)$ vs. $x/d$ data points while sampling only a small portion of the available $x$-$t$ data. As an example of automated analysis, a kymogram of 1,024 pixels in $x$ and 12,288 pixels in $t$ was analyzed by *Orientations* 1.1.x using a 50-pixel window size, yielding 4,760 $x$-$t$ slope measurements over the entire 2 s of real-time schlieren imagery. The minima of these data at 26 $x/d$ locations are plotted in Fig. 14 (yellow triangular symbols). The processing time required for this by the *Orientations* plugin in ImageJ was about 20 s on a typical desktop computer.

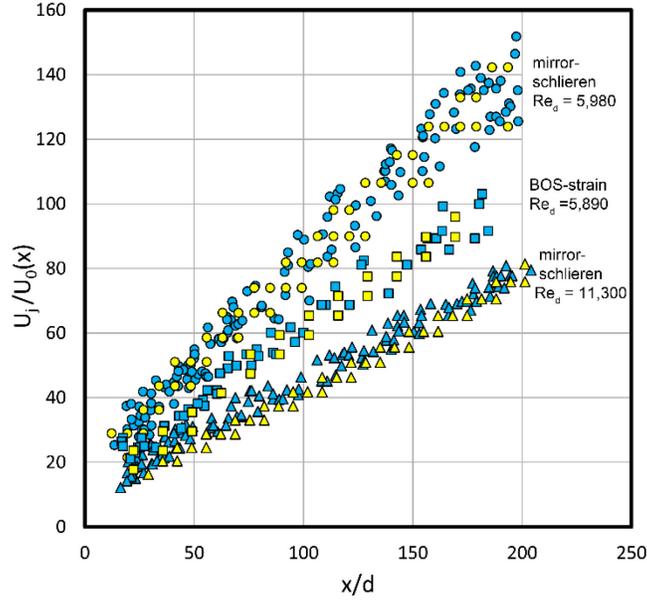

Fig. 14 Kymography summary graph of $U_j/U_0(x)$ vs. $x/d$ for three present test conditions, showing manually-reduced data by blue-filled (dark) symbols and ImageJ *Orientations* 1.1.x results by yellow-filled (light) symbols.

Finally, schlieren image stacks like those obtained here for jet velocimetry also can be interrogated digitally in various other ways, one of which is to compute the standard deviation $\sigma$ of the gray levels at each pixel in the stack and project the results onto a new 2D standard-deviation image [49, 50]. For present purposes such an image of the helium jet emphasizes regions of high turbulence intensity, as shown in Fig. 15a for $Re_d$ = 5,890, while remaining black where no turbulence is present. The standard deviation is highest for $10 \leq x/d \leq 60$, and is symmetrical about the jet centerline with a peak between the centerline and the jet edge. This peak agrees qualitatively with the turbulent kinetic energy profile of a self-similar RTJ shown by Panchapakesan and Lumley [51].

    The same standard-deviation image is shown again in Fig. 15b, now in grayscale and inscribed with the jet centerline and lines indicating $r/r_{1/2} = \pm 1$ and $\pm 2$. As defined with regard to Fig. 1, the jet spreading rate equals the slope of the $r/r_{1/2} = 1$ line, which is here measured to be 0.96. So, despite the density difference between the helium jet and the surrounding atmosphere, the present jet spreading rate closely matches that of benchmark airjet experiments in the literature [51, 52]. Further, as shown e.g. by Pope [25], the boundary-layer momentum equation requires that a linear jet spreading rate and self-similarity go hand-in-hand.

    The $r/r_{1/2} = \pm 2$ lines in Fig. 15b mark the mean outer jet boundary, although some occasional intermittent turbulence can still be seen outside this line. In angular measure the jet spreads at $\pm 11°$, or with a total spreading angle of $22°$.

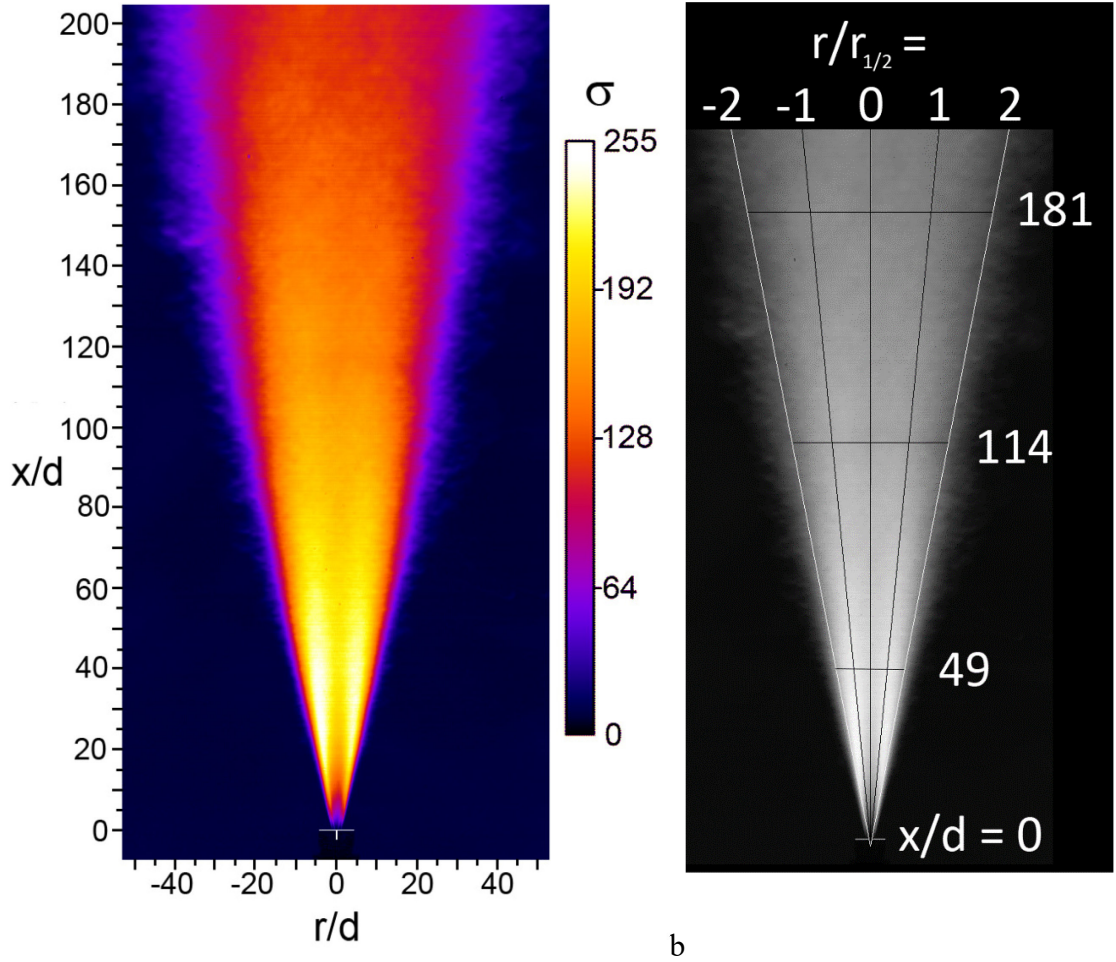

Fig. 15 Standard-deviation mirror-schlieren projection of a 12,288-image ensemble at $Re_d$ =5,890. a) The standard deviation $\sigma$ of gray levels from 0 to 255 is color-coded for clarity. b) Lines denoting $r/r_{1/2} = 0, \pm 1$, and $\pm 2$ are inscribed on the grayscale standard-deviation image.

## 5. Conclusions and Future Work

The current trend in schlieren optics is toward quantitative digital data acquisition and image processing, in addition to its earlier role of flow visualization [2, 27]. Seedless optical velocimetry of turbulent flows is an important component of this. BOS is another key component which was used until now only a few times for velocimetry, and which lacked a benchmark experiment for further development in that direction. Here, BOS and traditional mirror-schlieren imaging are compared for the first time in the velocimetry of a self-similar round turbulent jet, a canonical flow in turbulence research. Mirror-schlieren and BOS are each found to have advantages and disadvantages in this role.

The characteristics of the round turbulent jet are quite well known [25,28,41,51,52], but they are measured again in this paper to reveal how mirror-schlieren and BOS velocimetry behave when applied to such a well-established flowfield. Despite the previous success of schlieren velocimetry with 2D planar turbulent boundary layers, the round turbulent jet poses a serious challenge because the schlieren beam path-averages velocities across the jet [15]. Making sense of the results, without resorting to tomography, is found to depend primarily on the method of data analysis.

The Digital Image Correlation (DIC) is the traditional means of data analysis for PIV of seeded flows. Here, it is tasked with measuring the velocity of flows seeded only by naturally-occurring turbulence. This succeeded in previous schlieren velocimetry studies when the data were statistically adequate to capture flow intermittency [14-18]. For the present work we have acquired large image-data sets and demonstrated their statistical convergence.

The direct- or naïve-DIC analysis of these images gave poor velocimetry results, but the results were improved by gradient processing to extract strain-rate images before the velocimetry step. DIC was only expected to succeed for $x/d > 100$, where the camera frame rate is adequate compared to the speed of the jet, but even there it yields only a fraction of the actual jet velocity measured by traditional PIV. This behavior is recognized to be a consequence of velocity path-averaging across the axisymmetric jet.

Reasons for the increasingly-poor BOS DIC performance as $Re_d$ increases, and for the general failure of mirror-schlieren DIC velocimetry irrespective of $x/d$ or $Re_d$, are not so clear. The obtainable BOS resolution is limited, due to the required defocus of the jet with respect to the BOS background and the interrogation-window size in DIC processing. One might then expect a higher-$Re_d$ jet, having a broader range of turbulent scales, to be at a disadvantage for BOS, but this runs counter to our observation that the poor resolution of BOS still outperforms higher-resolution mirror-schlieren velocimetry by DIC processing. Mirror-schlieren possibly suffers from presenting naïve DIC with too much rapidly-changing information compared to BOS, the lack of a fixed background pattern, and the overlay of fast- and slow-moving turbulence that occurs in the schlieren image of an axisymmetric jet. It is also possible that the 50% mirror-schlieren knife-edge cutoff that we used was excessive for this purpose [14].

Compared to DIC, much better results were obtained by kymography (also known as streak-schlieren or $x$-$t$ diagram analysis). DIC is a pixel-wise analysis, while kymography is spatio-temporal. Also, DIC velocimetry is fully path-averaged, whereas kymography avoids path-averaging insofar as the jet centerline mean velocity $U_0(x)$ is concerned. Light rays passing through the jet centerline are still affected by off-centerline turbulence, but are shifted and distinguishable by their higher velocities in an $x$-$t$ diagram. In this unique case, the centerline velocity can be extracted without tomography: For our higher-$Re_d$ case and $x/d > 100$, the kymography results agree directly with our PIV benchmark mean-velocity data within the data scatter. Also, as reported by several past investigators [8, 15-18], we affirm that the eddy convective velocities measured in this experiment can sum to the mean-flow velocity, given a dataset large enough for statistical convergence.

Further, the kymograms presented here reveal that turbulent eddies last much longer than is required for them to be useful in schlieren velocimetry, contradicting one of its often-cited drawbacks. Kymography also succeeds at a lower frame rate than that required for traditional PIV, because the spacetime slope information of the turbulent schlieren streaks used for velocimetry is spread out across the kymogram in $x$ and $t$.

Overall, we found that the schlieren-derived jet centerline velocity $U_0(x)$ can be as low as 30% or as high as 100% of the true centerline velocity, depending upon the optical method, the method of extracting velocity data from schlieren images, and the jet Reynolds number $Re_d$. Reynolds number is not a primary scaling parameter of round turbulent jets [25], but there is a significant effect of $Re_d$ in the present schlieren velocimetry results (Figs. 11-13) that remains unexplained. Similar experiments at a higher $Re_d$ than tested here might help to resolve this issue.

Schlieren velocimetry confirms the self-similarity of the present round turbulent helium jet in air over a $Re_d$ range in terms of the self-similarity of transverse Gaussian velocity profiles, the expected $1/x$ centerline velocity decay, and the linear spreading rate of the jet. While this experiment thus provides a benchmark for schlieren velocimetry of the 2D axisymmetric mean turbulent jet flowfield, fully-3D flows still demand a more sophisticated tomographic approach. Nevertheless, what was learned here may be useful for schlieren tomography as well. For example, BOS DIC processing performed better in the present study, despite its low resolution, than mirror-schlieren velocimetry by DIC.

In summary, while traditional PIV is in several ways superior to schlieren velocimetry, useful results are nonetheless possible from the schlieren tracking of turbulence. Naïve DIC is not a good choice for schlieren velocimetry of axisymmetric turbulent flows like the present helium jet, but streak-schlieren kymography shows considerable promise, especially with automated rather than manual data reduction.

*Future Work* In addition to the higher jet $Re_d$ testing and discovering the mechanism by which schlieren velocimetry path-averages the jet velocity profile, mentioned earlier, we also suggest the following topics for future work:
- Other velocimetry approaches, such as optical flow and Wildeman's [30] Fourier algebraic demodulation method, which were beyond our scope, should also be tried using the helium jet in air as a testbed.
- Better spatial pixel resolution, a faster camera frame rate, and shorter exposure times are needed to study the nozzle near-field region of the helium jet, which was inadequately covered in the present work. A higher frame rate would also allow spectra of jet turbulence to be studied. (Spectra were limited to 3kHz frequency resolution by present frame rate.)
- The difficulty observed here with DIC velocimetry of mirror-schlieren images should be examined further by systematically varying the amount of schlieren knife-edge cutoff, and by performing mirror-shadowgraphy for comparison with schlieren imaging.

## Declaration of Competing Interests

The authors declare that they have no known competing financial interests or personal relationships that could have appeared to influence the work reported in this paper. This work received no external funding support.

## CRediT authorship contribution statement

**Gary Settles**: Conceptualization, Validation, Investigation, Writing –original draft, Writing - review & editing, Visualization. **Alex Liberzon**: Conceptualization, Software, Validation, Formal analysis, Writing –review & editing